\documentclass{mem}
\usepackage{natbib}\usepackage{txfonts}\usepackage{balance}
\usepackage{graphicx}
\usepackage[a4paper,breaklinks,dvipdfm]{hyperref}
\idline{75}{282}

\newcommand{\logg}{$\log g$}
\newcommand{\feh}{[Fe/H]}
\newcommand{\feone}{Fe\,{\sc I}}
\newcommand{\fetwo}{Fe\,{\sc II}}

\begin{document}
\def\teff{$T\rm_{eff }$}
\def\kms{$\mathrm {km s}^{-1}$}

\title{
Fundamental parameters of red giants in 47 Tucanae
}

   \subtitle{}

\author{
A. O. \, Thygesen\inst{1} 
\and L. \, Sbordone\inst{1,2}
          }

  \offprints{A. O. Thygesen}

\institute{
Zentrum f\"{u}r Astronomie der Universit\"{a}t Heidelberg -- Landessternwarte,
K\"{o}nigstuhl 12,
69117 Heidelberg, Germany.
\and
GEPI, Observatoire de Paris, CNRS, Univ. Paris Diderot, Place Jules Janssen, 92190, Meudon, France.
\email{a.thygesen@lsw.uni-heidelberg.de}
}

\authorrunning{Thygesen}

\titlerunning{47 Tucanae giant parameters}

\abstract{ In these proceedings we present initial results of a spectroscopic analysis of a small sample of evolved K-giants in the globular cluster 47 Tucanae. We derive the effective temperature, surface gravity and microturbulence of all targets using standard methods. Further we derive LTE abundances of [Fe/H] and [O/Fe] as well as the abundance of sodium, using NLTE corrections. We find a mean metallicity for the cluster of [Fe/H]$=-0.75\pm0.10$, in excellent agreement with several other studies. Also, we confirm the sodium-oxygen anticorrelation previously reported by a number of other authors. Finally, we see indications of the sodium enriched stars also being enriched in heavy magnesium isotopes. 

}
\maketitle{}

\section{Introduction}
Studying elemental abundances in globular clusters is a key to understand the chemical evolution of stellar populations, due to the confined environment in which the different stellar generations are born. By now it is well established that most, if not all, globular clusters show multiple stellar populations, both on the main sequence as well as on the red giant branch, indicating a more complex history than previously thought (see e.g. \citealt{carretta2}). The multiple populations are visible in e.g. different chemical abundance patterns and different photometric colors. This is believed to be caused by cluster self-pollution from a previous generation of stars, but the exact nature of these stars is still debated. Recently, multiple sequences were discovered in 47 Tucanae \citep{milone}, one of the most extensively studied globulars. We have analysed high resolution (R=110000) high signal-to-noise ($\geq160$) spectra of a number of evolved K-giants in this cluster obtained with the VLT UVES spectrograph. From these we determine the effective temperature, \teff, surface gravity, \logg, microturbulence, $\xi_t$, and derive elemental abundances of O, Na and Fe to investigate the Na-O anticorrelation observed in all globular clusters. The abundances have been derived from standard equivalent width measurements. This work is part of the initial analysis to derive accurate Mg-isotopic ratios for all stars in our 47 Tucanae sample.

\section{Spectroscopic analysis}
From the high-resolution spectra we performed a standard spectroscopic analysis. The atomic lines were carefully selected to have reliable atomic data as well as being reasonably free of strong blends. For all relevant lines we manually measured equivalent widths (EW) using the standard IRAF {\tt splot} task, including deblending when needed. The hereby measured EW's were subsequently analyzed with the spectral analysis package MOOG\footnote{available from http://www.as.utexas.edu/chris/moog.html}. The stellar parameters were derived in the standard way of requiring no correlation of iron abundance vs. EW ($\xi_t$), abundance vs. excitation potential (\teff) and ionization equilibrium between \feone\ and \fetwo\ (\logg). 

We employed interpolated model atmospheres from a grid of $\alpha$-enhanced 1D ATLAS9 atmospheric models in the analysis \citep{castelli}. Using the best-fitting model, the abundance of sodium was determined by fitting the measured EW's. For the Na lines we applied NLTE corrections on a line-by-line basis, using the values from \citet{lind}. The corrections were based on our derived stellar parameters, or the closest match in the grid calculated by Lind et al. if no exact match was available\footnote{see http://inspect-stars.net/}. After applying the corrections, a mean abundance was derived. To measure the oxygen abundance we performed spectral synthesis (again using MOOG) of the forbidden transition O-line at 6300Å as well as the O-line at 6363Å. However, the latter could not be synthesized succesfully in all cases due to the presence of atmospheric emission lines. For one target we could only put a lower limit on the oxygen abundance, due to emission lines being present in both available lines (marked with an arrow in Fig.~\ref{nao}). The abundances of all elements were measured relative to the Sun, using the \citet{lodders} solar abundances, but including the recent updates from \citet{caffau}.

\section{Results}
\begin{figure}[htb]	
\centering
\includegraphics[width=\columnwidth]{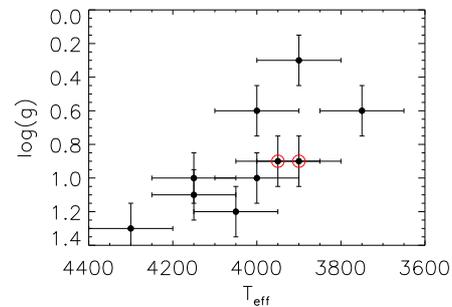}
\caption{Basic parameters of the giants in the sample investigated in this work. Two giants with very similar parameters are indicated with red circles (See Figs.~\ref{nao} and ~\ref{mgh})}
\label{cmd}
\end{figure}

      In Fig.~\ref{cmd} is shown the position of our giants in a \logg\ vs. \teff\ diagram and as can be seen, we sample a set of very evolved giants, with the most extreme case being a potential AGB star. We find good agreement between our spectroscopic temperatures and those derived from broadband photometry, using the calibration of \citet{ramirez}, with a deviation of only $40\pm86$ K. From our [Fe/H] measurements we derive a mean metallicity for the cluster giants of [Fe/H]$ =-0.75\pm0.10 $ which is in excellent agreement with the recent study of \citet{koch}, as well as several other studies of 47 Tuc (e.g. \citealt{carretta2,alves-brito}).
      
\begin{figure}%
\centering
\includegraphics[width=\columnwidth]{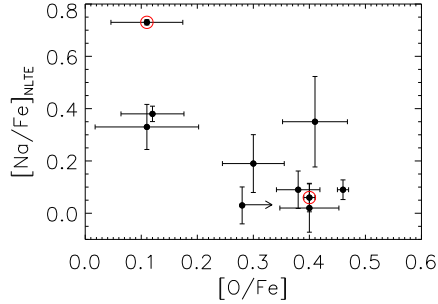}%
\caption{The anti-correlation of Na and O in our sample. As can be seen, the stars appear to separate in two groups, potentially indicating membership of two different populations.}%
\label{nao}%
\end{figure}

      In Fig.~\ref{nao} we present the abundances of sodium and oxygen, relative to the iron abundance of the stars. The Na-O anticorrelation found in all globular clusters is clearly present in our sample of giants from 47 Tuc, confirming the results of \citet{carretta} and Koch \& McWilliam, where the latter saw an indication of this in their sample of 9 giants. However, the anticorrelation is significantly more pronounced in our sample, compared to Koch \& McWilliam. We note that the Na-O anticorrelation has also been observed in a sample of dwarf stars in 47 Tuc \citep{dorazi} showing that this is a general property of the cluster, not related to the giant nature of the stars in our work. Interestingly, we see an indication that the giants may fall in two discrete subgroups, rather than along a continuous distribution of Na/O ratios, with one being strongly enriched in Na compared to the O-enriched stars. However, a larger sample will need to be studied before any firm conclusions can be made.
      
\begin{figure}%
\centering
\includegraphics[width=\columnwidth]{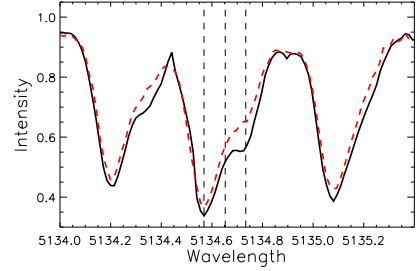}%
\caption{Plot of the observed spectra of two giants from our dataset with identical \logg, and \teff\ only differing by 50K. Both stars are indicated in Figs.~\ref{cmd} and \ref{nao}. The star plotted with a black line is strongly enriched in Na and depleted in O. The opposite is the case for the star shown with the red dashed line. Shown is the region around one of the MgH bands typically used for isotopic measurements. The line-centers of $^{24}$MgH, $^{25}$MgH and $^{26}$MgH are indicated with vertical, dashed lines (left to right).}%
\label{mgh}%
\end{figure}

\begin{figure*}%
\centering
\includegraphics[width=0.95\textwidth]{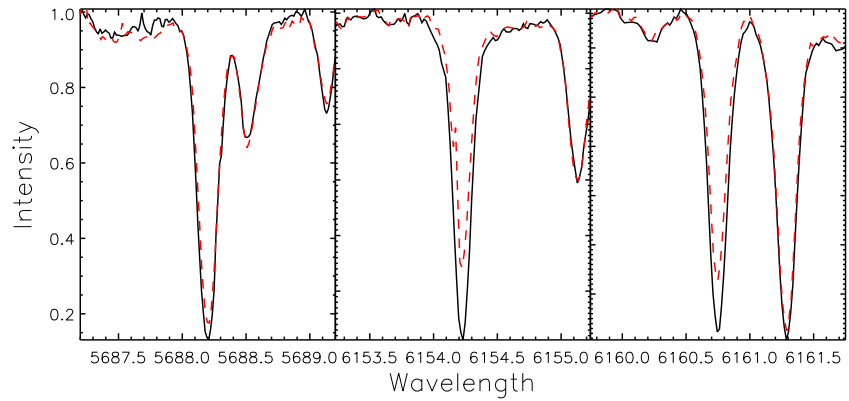}
\caption{The Na-lines for the same two stars as in Fig.~\ref{mgh}. The difference in sodium abundance is evident.}%
\label{nalines}%
\end{figure*}

      The final goal of this project is the study of magnesium isotopic ratios from the MgH bands around 5130Å, so it is natural to look for differences in the molecular bands simply by visually inspecting the observed spectra. In Fig.~\ref{mgh} is shown spectra of two giants with almost identical parameters (identical \logg\, 50K difference in \teff, marked with red in Fig.~\ref{cmd} and \ref{nao}), but one being strongly enriched in Na and O depleted (black line), while the other is O-rich and Na-poor (red dashed line, see also Fig.~\ref{nalines}). As is immediately obvious, the Na-enriched star is also strongly enhanced in the heavy magnesium isotopes. If this is the case also for the remaining sample of Na-enriched stars this implies that the nucleosynthesis of heavy Mg isotopes takes place in the same kind of polluters who produce the enrichment of sodium.

\section{Conclusions}
 We have determined the fundamental parameters for a sample of giants in the globular cluster 47 Tucanae. Using 1D atmospheric models, we derive the abundances of Fe, Na and O, confirming both the metallicity of the cluster (\feh$=-0.75\pm0.10$) as well as the Na-O anticorrelation found by several other authors. It is found that our targets appear to separate into two groups, one being Na enriched and oxygen depleted, whereas the other is found to be rich in oxygen, but with a much lower sodium abundance. Visually inspecting spectra of two representative stars, one from each sample, we see a clear difference in the enrichment of heavy magnesium isotopes. As the Na-enriched giant show much higher abundances of the heavy magnesium isotopes this indicates that they are created in the same polluters who are responsible for creating the overabundance of Na. However, detailed study of the Mg istopic ratios for all stars in the sample will be needed to quantify this further.

\begin{acknowledgements}
AOT acknowledges support from Sonderforschungsbereich SFB 881 "The Milky Way System" (subproject A5) of the German Research Foundation (DFG).
\end{acknowledgements}

\bibliographystyle{aa}
\bibliography{rome-proceedings}

\end{document}